\documentclass[%
 preprint,
superscriptaddress,
 amsmath,amssymb,
 aip,
floatfix,
longbibliography
]{revtex4-2}
\usepackage[utf8]{inputenc}
\usepackage{graphicx}
\graphicspath{{Figures/}}
\usepackage{epstopdf}
\usepackage{amsmath,amssymb}
\usepackage{xcolor}
\usepackage{graphicx}
\usepackage{dcolumn}
\usepackage{bm}
\usepackage{hyperref}

\begin{document}
\preprint{AIP/123-QED}

\title{Canard explosions in turbulent thermo-fluid systems}

\author{Ramesh S. Bhavi}
\email{rameshbhavi003@gmail.com}
\author{Sivakumar Sudarsanan}%
\author{Manikandan Raghunathan}
\author{Anaswara Bhaskaran}
\author{R. I. Sujith}
\affiliation{Department of Aerospace Engineering, Indian Institute of Technology Madras, Chennai, Tamil Nadu 600036, India.}
\affiliation{Centre of Excellence for Studying Critical Transition in Complex Systems, Indian Institute of Technology Madras, Chennai, Tamil Nadu 600036, India.}
\date{\today}
\begin{abstract}
 A sudden transition to a state of high amplitude limit cycle oscillations is catastrophic in a thermo-fluid system. Conventionally, upon varying the control parameter, a sudden transition is observed as an abrupt jump in the amplitude of the fluctuations in these systems. In contrast, we present an experimental discovery of a canard explosion in a turbulent reactive flow system where we observe a continuous bifurcation with a rapid rise in the amplitude of the fluctuations within a narrow range of control parameters. The observed transition is facilitated via a state of bursting, consisting of the epochs of large amplitude periodic oscillations amidst the epochs of low amplitude periodic oscillations. The amplitude of the bursts is higher than the amplitude of the bursts of intermittency state in a conventional gradual transition, as reported in turbulent reactive flow systems. During the bursting state, we observe that temperature fluctuations of exhaust gas vary at a slower time scale in correlation with the amplitude envelope of the bursts. We also present a phenomenological model for thermoacoustic systems to describe the observed canard explosion. Using the model, we explain that the large amplitude bursts occur due to the slow-fast dynamics at the bifurcation regime of the canard explosion.
 
\end{abstract}

\maketitle

\begin{quotation}
Transition to oscillatory instabilities in turbulent reactive flow systems is a long pending issue in designing modern combustors of engines that have high-power ratings. Nonlinear interactions between the hydrodynamic flow field, the acoustic field and the heat-release rate fluctuations in a confined environment make a turbulent combustor a complex dynamical system. The state of these systems changes from a stable operation to a state of oscillatory instability as the control parameter is varied. In turbulent combustors, past studies were focused on the gradual transitions to the state of oscillatory instability via the state of intermittency. Most recently, the discovery of abrupt transitions in turbulent reactive flow systems has been a highlight, which is a contrasting scenario of a gradual transition. Abrupt transitions are sudden and discontinuous in nature. However, in this study, we report the discovery of a canard explosion, which is a transition involving a rapid rise in the amplitude of the oscillations but continuous in nature. Canard explosions are characterized by the amplitude of the oscillations reaching a very high value within a narrow range of control parameters. Further, we also observe that the transition is facilitated via the state of bursting, where the bursts are of large amplitude. We show that such bursts are possible when there is a fluctuation in the parameter at the bifurcation regime of the underlying canard explosion.


\end{quotation}

\section{Introduction} \label{Sec: Introduction}

Emergent oscillatory instabilities are well-known in fluid mechanical systems. Such instabilities are observed in thermoacoustic \cite{sujith2021thermoacoustic}, aeroacoustic \cite{hirschberg2004introduction}, and aeroelastic systems \cite{hansen2007aeroelastic}. The state of oscillatory instabilities corresponds to the unstable operation in many such systems. The large amplitude oscillations during the state of instability hamper healthy working conditions of these engineering systems, consequently leading to catastrophic failures \cite{parkinson1971wind,lieuwen1999investigation,brown1981vortex}. These oscillatory instabilities arise due to the nonlinear interactions between the sub-systems of a fluid mechanical system, as a control parameter is changed.

The transition from a stable operation to an unstable operation in a dynamical system is referred to as a bifurcation to the state of limit cycle oscillations (LCO) \cite{lieuwen2002experimental,strogatz2018nonlinear}. In laminar thermo-fluid systems, where the dynamics of the flow is calm and quiet, the transition is a Hopf-bifurcation as the system transits from a silent state (fixed point) to an oscillatory state \cite{etikyala2017change,subramanian2010bifurcation}. In the case of turbulent systems, the dynamics comprise vigorous turbulent fluctuations in the flow. In these turbulent systems, the stable operation is characterized by chaotic oscillations \cite{gotoda2011dynamic}, and the unstable operation corresponds to an ordered state of periodic oscillations \cite{mondal2017onset}. Studies in the recent decade have shown that the state of intermittency, an asymptotic state which has the imprints of chaos and order, presages the emergence of order \cite{nair2014intermittency}. The emergence of order via the state of intermittency is predominantly observed as a gradual change in the root mean square (RMS) value, a statistical measure of acoustic pressure oscillations. Hence, in turbulent systems, the bifurcation is viewed as a gradual emergence of order from the state of chaos \cite{mondal2017onset,pavithran2020universality}. In contrast to the gradual transition via the state of intermittency, recently, abrupt transitions to the state of order have also been discovered in turbulent systems \cite{singh2021intermittency,bhavi2023abrupttransition,joseph2024explosive}. In abrupt transition, a sudden discontinuous jump in the RMS of the acoustic pressure oscillations is observed.

Abrupt transitions are also referred to as explosive transitions and are characterized by the phenomenon of hysteresis \cite{kumar2015experimental}. The occurrence of hysteresis is due to the simultaneous presence of multiple stable regimes for a range of control parameters \cite{zou2014basin,bhavi2023abrupttransition}. However, in practical engineering systems, there are exceptions where a genuine abrupt rise in the statistical measure of the oscillations is observed, but the transition is not discontinuous \cite{brons1991canard}. Such transitions, where a rapid rise in the amplitude of the fluctuation occurs for a minute increment in the control parameter, were primarily investigated in the Van der Pol oscillator model and are referred to as canard explosions \cite{krupa2001relaxation}. Canard explosions have been reported in many real-world systems such as chemical oscillations \cite{brons1991canard}, ground dynamics of an aircraft \cite{rankin2011canard}, neuronal activity \cite{moehlis2006canards}, predator-prey food chains \cite{deng2004food}, and light emitting diodes \cite{marino2011mixed}.

In a transition involving a canard explosion, the amplitude of the limit cycle grows significantly soon after the Hopf bifurcation \cite{borgers2017introduction}. The dynamics of the system during a canard explosion becomes highly sensitive to variation in the control parameter. There is a significant growth in the amplitude of the oscillation for an exponentially small range of values of the control parameter at the canard explosion regime \cite{brons1991canard}. Hence, a canard explosion appears abrupt if there is a lack of resolution in the variation in system parameters \cite{diener1984canard}. A continuous transition comprising a canard explosion, albeit appears abrupt, traces the same forward and reverse path in the control parameter variation \cite{borgers2017introduction}. Further, large amplitude bursts and mixed-mode oscillations are observed when the system exhibits slow-fast dynamics at the canard explosion regime \cite{han2012slow,desroches2013mixed}.

Here, we report the observation of canard explosions in thermo-fluid systems for the first time, to the best of our knowledge. We present the experimental results for the rapid rise in amplitude of the acoustic pressure oscillations within a minute range of the control parameter, a principal feature of the canard explosion. The transition is continuous in nature and exhibits no hysteresis. We also observe a bursting behaviour comprising the bursts of large amplitude acoustic pressure oscillations near the canard explosion regime. Through experimentally measuring the exhaust gas temperature during the state of bursting, we show that a system parameter fluctuates at a time scale slower than the system oscillations. Further, we describe the observed transition of canard explosion using a low-order thermoacoustic model. Using the low-order model, we attribute the bursting behaviour during the canard explosion to a coupling between a slow oscillatory term and the driving term. 

The rest of the paper is organized as follows: Section \ref{Sec: Experimental setup} provides a detailed description of the experiments and the setups used in this study. The experimental results of the transitions involving canard explosions are described in Section \ref{Sec: Experimental results for canard explosions}. The low-order thermoacoustic model describing canard explosions is presented in Section \ref{Sec: Thermoacoustic model of Canard explosio}, and the mechanism of large amplitude bursting dynamics is illustrated in Section \ref{Sec: Illustrate bursts using model}. Section \ref{Sec: Conclusions} narrates the conclusions of the study.

\section{Experiments}\label{Sec: Experimental setup}
In order to check the commonality of the transition to oscillatory instabilities in different turbulent reactive flow systems, we conducted experiments in three different configurations of combustors. These systems function in turbulent conditions and represent the dynamics of combustors in modern gas turbines and rocket engines. The details of the combustor setups are discussed below.

\subsection{ Dump combustor configurations}
\label{Sec: Dump Combustors}
 Figure~\ref{experimentalsetups}(a) represents the experimental setup for the dump combustor. A fluid mixture of compressed air and liquid petroleum gas (60\% Propane \& 40\% butane) is used for chemical reactions in a combustion chamber. The combustion chamber is 1100 mm long and has a 90 $\times$ 90 $\mathrm{mm}^2$ square cross-section. The setup has three main sections along the fluid flow--- a plenum chamber, a burner, and the combustion chamber. The air enters the combustor via a flow equalization chamber referred to as a plenum chamber, which helps isolate the combustion chamber from the fluctuations upstream of the flow. The fuel is injected in the burner section between the plenum chamber and the combustion chamber, where the fuel and the air are premixed. The diameter of the burner is 40 mm. The fuel-air mixture enters the combustion chamber at the dump plane, where there is a sudden increase in the cross-sectional area from the burner to the combustion chamber. The exit of the combustion chamber is connected to a large rectangular box referred to as a decoupler. The dimensions of the decoupler are set to be much larger than the cross-sectional dimensions of the combustion chamber. The utility of the decoupler is to reduce sound emissions from the combustion chamber \cite{Zinn1996}.
 
 The dynamics of the system is studied by varying the equivalence ratio $\phi$ as the control parameter. The equivalence ratio is defined as $\phi = \Upsilon_\mathrm{actual} / \Upsilon_\mathrm{stoichiometric}$, where  $\Upsilon$ is the ratio of the mass flow rate of the fuel and the air. Thus, $\phi$ is a function of air and fuel flow rates, which are controlled using mass flow controllers (MFC). The uncertainty in the flow rate measurement is $\pm$(0.8 $\%$ of the reading + 0.2 $\%$ of the full scale).  The uncertainty in the computed value of $\phi$ is $\pm 2\%$. The control parameter ($\phi$) is varied in a quasi-static manner. The qualitative change in the behaviour of the system is analyzed by measuring the acoustic pressure fluctuations in the combustion chamber. We used Piezoelectric pressure transducers (PCB103B02) for measuring the acoustic field fluctuations. The sensitivity of the transducers is 217.5 mV/kPa. We acquire the signal from the pressure transducer for 5 s at a sampling rate of 10 kHz after an initial waiting time of 3 s at each set point of the control parameter. The maximum uncertainty in the measured values of the pressure signal is $\pm 0.15$ Pa. The experiments were performed in two different configurations of the dump combustor, which will be detailed in the following subsections.

\subsubsection{ Dump combustor with a swirler configuration}
\label{Sec: Swirl stabilized dump Combustor}
 A swirler (refer to Fig.~\ref{experimentalsetups}b), inducing swirl motion to the flow, is used at the entry of the combustion chamber. The swirling motion aids in the establishment of the flame in a compact form, stretching over a small section of the combustion chamber. The diameter (d) of the swirler is $40$ mm. The swirler consists of 8 vanes, with each vane having an angle of $40^{\circ}$ with respect to the direction of the bulk blow in the combustor. The location of the swirler is such that the front part of each vane is 20 mm from the dump plane. In this swirler configuration, we maintain a constant fuel flow rate of 28 standard litres per minute (SLPM). The equivalence ratio varies from 0.783 to 0.532 by increasing the airflow rate from 800 SLPM to 1436 SLPM. The Reynolds number for the system, based on the diameter of the swirler, changes between $Re_d = 2.76 \times 10^4 \pm 220$ and $4.94 \times 10^4 \pm 220$. A K-type thermocouple is used to measure the temperature of the hot gases downstream of the flow. The signal for the temperature was acquired for 5 s at a sampling rate of 20 Hz. 

 \subsubsection{Dump combustor with a bluff body configuration}
\label{Sec: Bluff body stabilized dump Combustor}
\begin{figure}[h]
\centering
\includegraphics[width=0.5\linewidth]{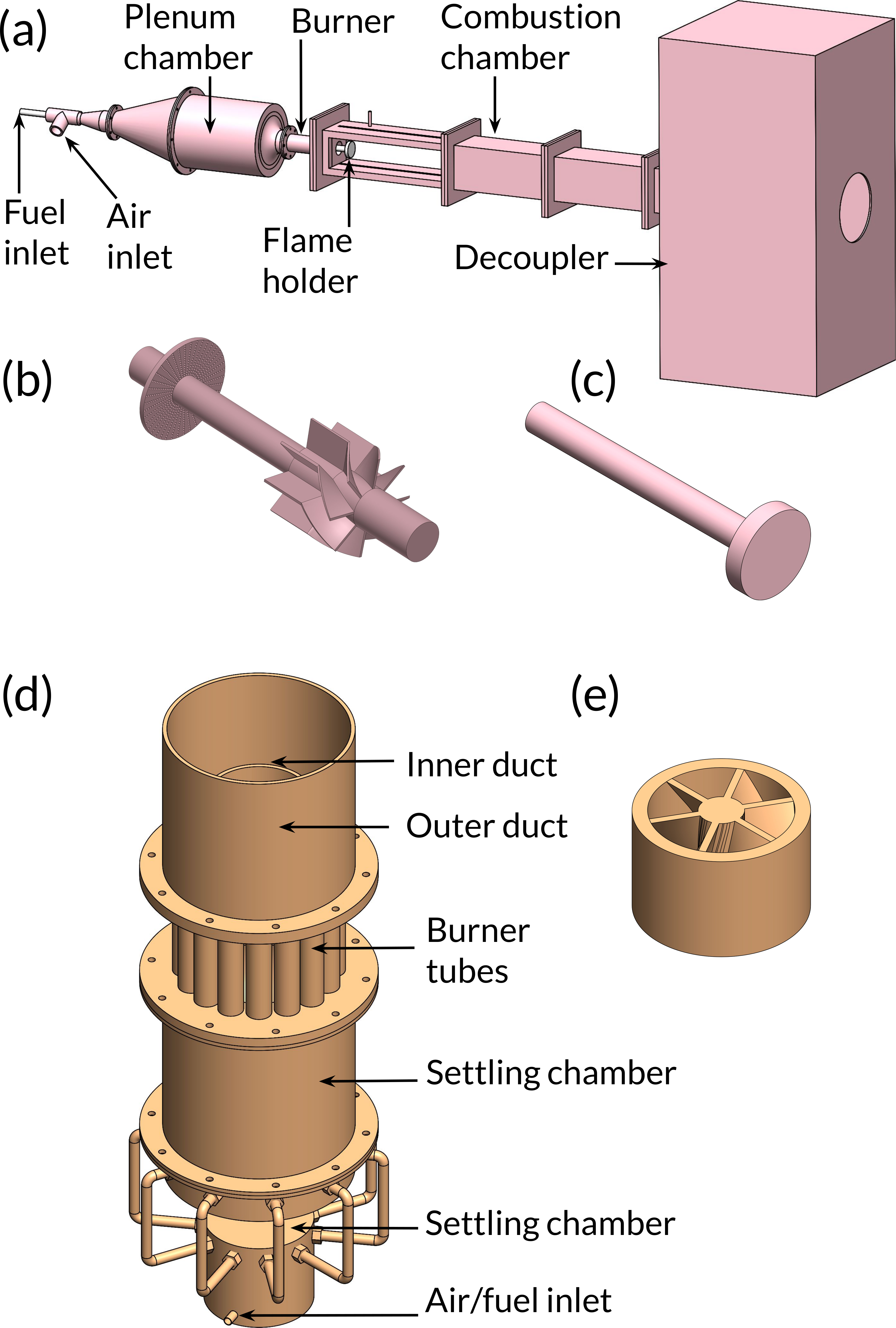}
\caption{Schematic of \textbf{(a)} a dump combustor which can be operated in two different configurations based on the flame holding mechanisms. We use \textbf{(b)} the swirler and \textbf{(c)} the bluff body as two different flame holders for the dump combustor. Schematic of \textbf{(d)} the annular combustor comprising sixteen burners. At the exit of each burner, \textbf{(e)} a swirler is used as a flame holder.}
    \label{experimentalsetups}
\end{figure}

 In this configuration of the dump combustor, we replace the earlier flame holder (swirler) with a bluff body (refer to \ref{experimentalsetups}c). A bluff body slows the flow by creating a flow re-circulation zone, providing sufficient time for the air-fuel mixture to react in a compact zone of the combustion chamber \cite{chen1990studies}. The bluff body is located at a distance of 27.5 mm from the dump plane of the combustion chamber. The diameter (d) of the bluff body is $47$ mm. The fuel for the combustor is introduced in the burner at a distance of 85 mm from the dump plane, through the hollow shaft anchoring the bluff body. We maintain a constant fuel flow rate of 42 SLPM in this bluff body configuration. The equivalence ratio varies from 1.909 to 1.022 by increasing the airflow rate from 600 SLPM to 1200 SLPM. The corresponding Reynolds number, computed based on the diameter of the bluff body, changes in the range of $Re_d = 1.76 \times 10^4 \pm 220$ to $3.28 \times 10^4 \pm 220$.

\subsection{Annular combustor}
\label{Sec: Annular combustor setup}
Figure~\ref{experimentalsetups}(d) represents a swirl-stabilized annular combustor, where sixteen flames from the circumferentially arranged burners are established during the experiments. Premixed air and LPG are used for chemical reactions. The air and the fuel initially enter a premixing chamber through an air/fuel inlet. The premixed mixture then enters into a flow-settling chamber. We incorporate a honeycomb-like structure inside the settling chamber to render the flow in one direction. The flow through the settling chamber encounters a hemispherical flow divider that uniformly distributes the fuel-air mixtures to the 16 burner tubes. The burner tubes exit into the combustion chamber comprising an outer and inner cylindrical duct. The chemical reactions are individually established in the annulus of the outer and the inner cylindrical duct after passing through the swirler fitted at the exit of each burner tube. The swirlers consist of vanes which are inclined at an angle of $\beta = 60^{\circ}$ with the axial flow direction (refer to Fig.~\ref{experimentalsetups}e). The burner tubes are 300 mm long and have a circular cross-section (30 mm diameter). The diameter of the inner and the outer cylindrical ducts are 400 mm and 300 mm, respectively. The length of the inner and the outer cylindrical ducts are 510 mm and 140 mm, respectively.
 
The equivalence ratio ($\phi$) is varied from 1.4 to 0.9 in a quasi-static manner by varying the fuel flow rate. The airflow rate is kept constant at 1800 SLPM throughout the experiments. The fuel flow rate is varied from 92 to 59 SLPM. The Reynolds number, calculated using the exit diameter of the burner, is $Re_d \approx 1.01 \times 10^4 \pm 220$. The dynamics of the system is analyzed by measuring the acoustic pressure fluctuations from the combustion chamber. Piezoelectric pressure transducers (PCB103B02) of sensitivity 217.5 mV/kPa are used for pressure fluctuation measurements. The pressure signal at each control parameter is acquired for 5 s at a sampling rate of 10 kHz after an initial waiting time of 3 s at each set point of the control parameter. A K-type thermocouple is used to measure the temperature of the hot gases downstream of the flow.

\section{Canard explosions in turbulent combustors}\label{Sec: Experimental results for canard explosions}

\begin{figure*}
\centering
\includegraphics[width=0.9\linewidth]{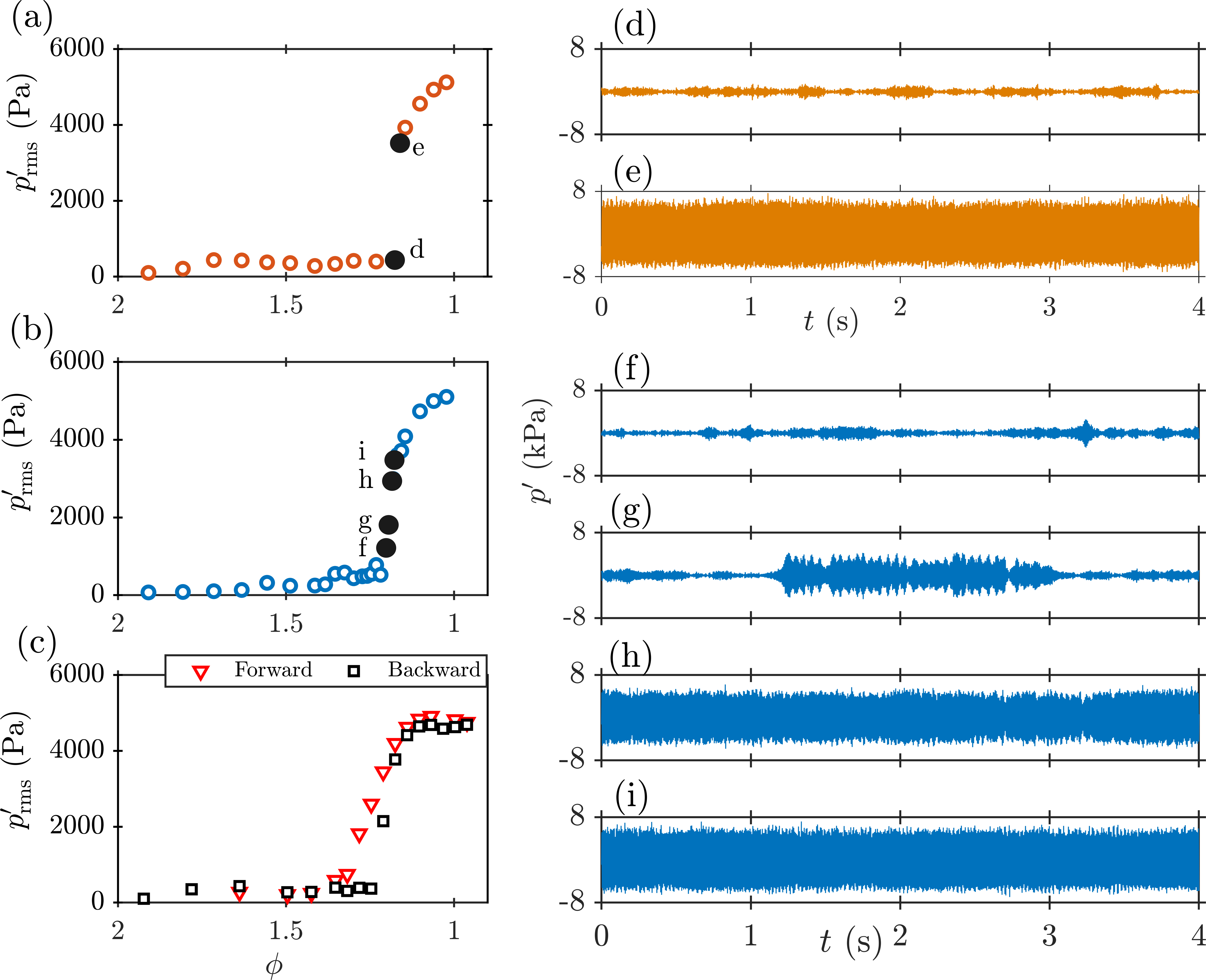}
\caption{Representation of a sudden transition to high amplitude limit cycle oscillations via canard explosion in the bluff body stabilized dump combustor. (\textbf{a,b \& c}) The bifurcation diagrams for the variation of the RMS value of the acoustic pressure fluctuations ($p'_\mathrm{rms}$) as a function of the equivalence ratio $\phi$. (\textbf{d-i}) The corresponding time series of the acoustic pressure signal during canard explosion.\textbf{(a)} Sudden transition from a low amplitude \textbf{(d)} ($p'_\mathrm{rms} = 420$ Pa) to very high amplitude \textbf{(e)} ($p'_\mathrm{rms}$ = 3525 Pa) acoustic pressure fluctuations as $\phi$ is varied. When $\phi$ is varied in finer steps between these apparently abrupt transition points, we observe a \textbf{(b)} continuous transition \textbf{(f-i)} to high amplitude fluctuations via \textbf{(g)} bursting dynamics. \textbf{(c)} The transition retraces the forward path when $\phi$ is varied in the reverse direction, implying the absence of the hysteresis. Thus, we observe a transition with a rapid rise in the amplitude of acoustic pressure fluctuations at a certain value of the control parameter.}
    \label{Fig: Bifurcation bluff body}
\end{figure*}

Figure~\ref{Fig: Bifurcation bluff body} represents the bifurcation diagram and the nature of the sudden transition in the bluff body stabilized dump combustor. In order to study the sudden transitions via canard explosions, we varied the equivalence ratio ($\phi$) as the control parameter. Initially, when the airflow rate is varied in steps of 30 SLPM, we observed an abrupt transition (refer to points d and e in Fig.~\ref{Fig: Bifurcation bluff body}a). The abrupt transition is from low amplitude ($p'_\mathrm{rms}$ = 420 Pa) to high amplitude ($p'_\mathrm{rms}$ = 3525 Pa) acoustic pressure fluctuations. Here, $p'_\mathrm{rms}$ represents the root mean square value of the acoustic pressure fluctuations ($p'$). The corresponding time series are presented in Fig.~\ref{Fig: Bifurcation bluff body}(d, e). To further investigate this seemingly abrupt transition, we varied the airflow rate at finer steps (10 SLPM) between the points of the control parameter corresponding to the abrupt jump. A continuous, albeit steep, variation in the RMS value of the $p'$ is observed when the control parameter is varied in finer steps (refer to Fig.~\ref{Fig: Bifurcation bluff body}b). We further note that the continuous transition occurs via a state of bursting (refer to Fig.~\ref{Fig: Bifurcation bluff body}f-i). During the state of bursting, we observe large amplitude fluctuations ($p' \approx 3500$ Pa) amidst low amplitude fluctuations ($p' \approx 500$ Pa) (refer to Fig.~\ref{Fig: Bifurcation bluff body}g).

\begin{figure} 
\centering
\includegraphics[scale=0.74]{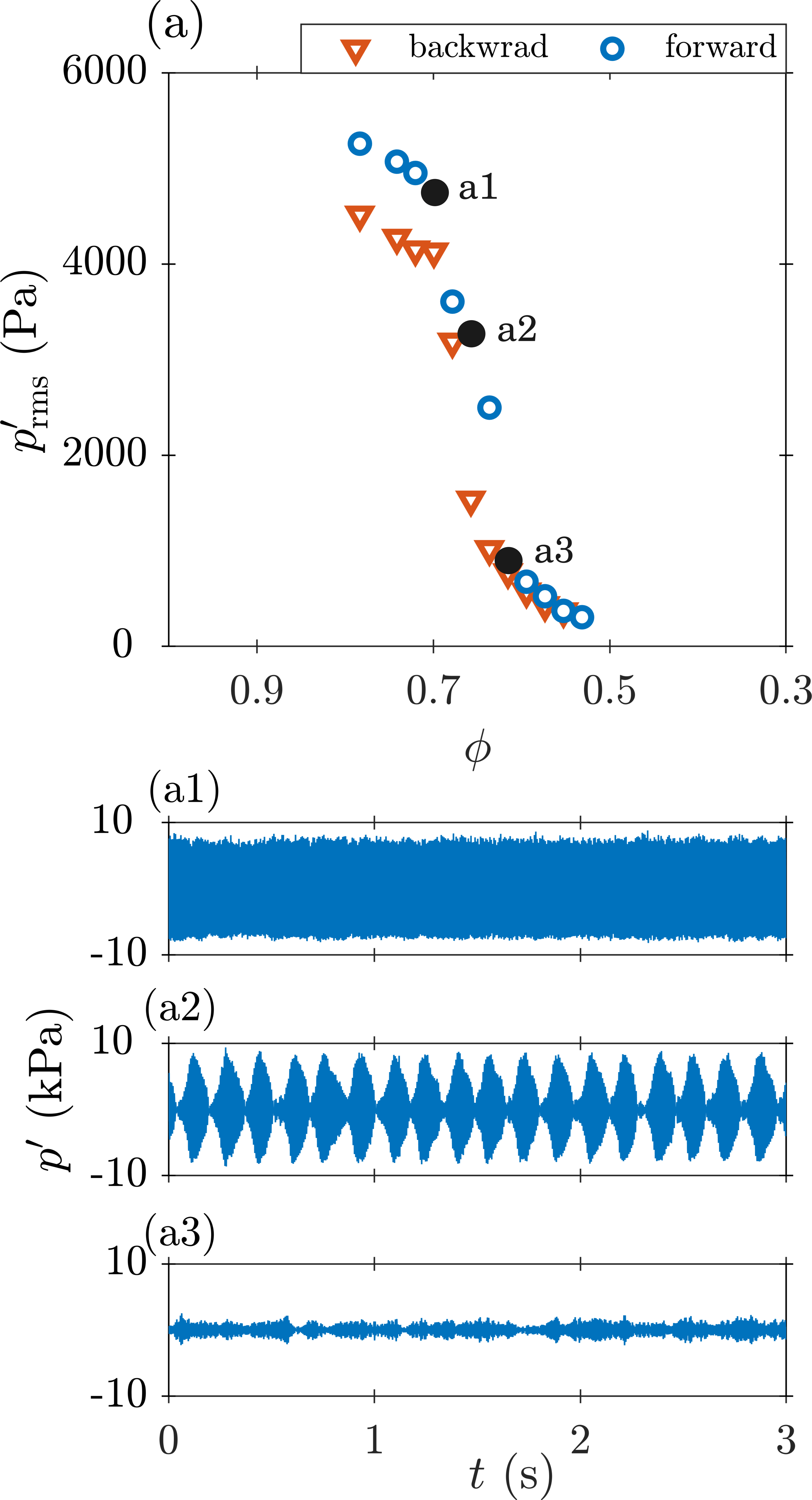}
\caption{Representation of the sudden transition via canard explosion in a swirl stabilized dump combustor. \textbf{(a)} The forward and reverse path of the transition via the canard explosion. The equivalence ratio ($\phi$) in the forward path is varied from 0.783 to 0.532. We notice that the transition occurs via the state of large amplitude bursting \textbf{(a1-a3)}.}
    \label{Fig: Bifurcation swirler}
\end{figure}

Further, when the control parameter is varied in the reverse direction, the transition retraces the forward path (refer to Fig.~\ref{Fig: Bifurcation bluff body}c). Similar observations of the canard explosions were observed when we performed experiments in a swirl-stabilized dump combustor (Fig~\ref{Fig: Bifurcation swirler}). In the swirl-stabilized dump combustor, as the equivalence ratio is decreased from 0.783 to 0.532, we observe a rapid decrease in the variation of RMS value of the acoustic pressure fluctuations (refer to the points a1, a2, \& a3 of Fig.~\ref{Fig: Bifurcation swirler}a). The transition is from a state of high amplitude fluctuations ($p'_\mathrm{rms}$ = 4730 Pa) to a state of low amplitude fluctuations ($p'_\mathrm{rms}$ = 770 Pa) (refer to Fig.~\ref{Fig: Bifurcation swirler}a1, a3). Additionally, we note that when the parameter is varied in the reverse direction, the system retraces the forward path (Fig.~\ref{Fig: Bifurcation swirler}a). The difference in the values of $p'_\mathrm{rms}$ at the state of thermoacoustic instability in forward and reverse paths is due to increased damping as a result of prolonged heating of the combustor walls \cite{pavithran2023tipping}. Thus, we note that a continuous but steep transition involving a canard explosion exhibits no hysteresis.

\begin{figure} 
\centering
\includegraphics[scale=0.74]{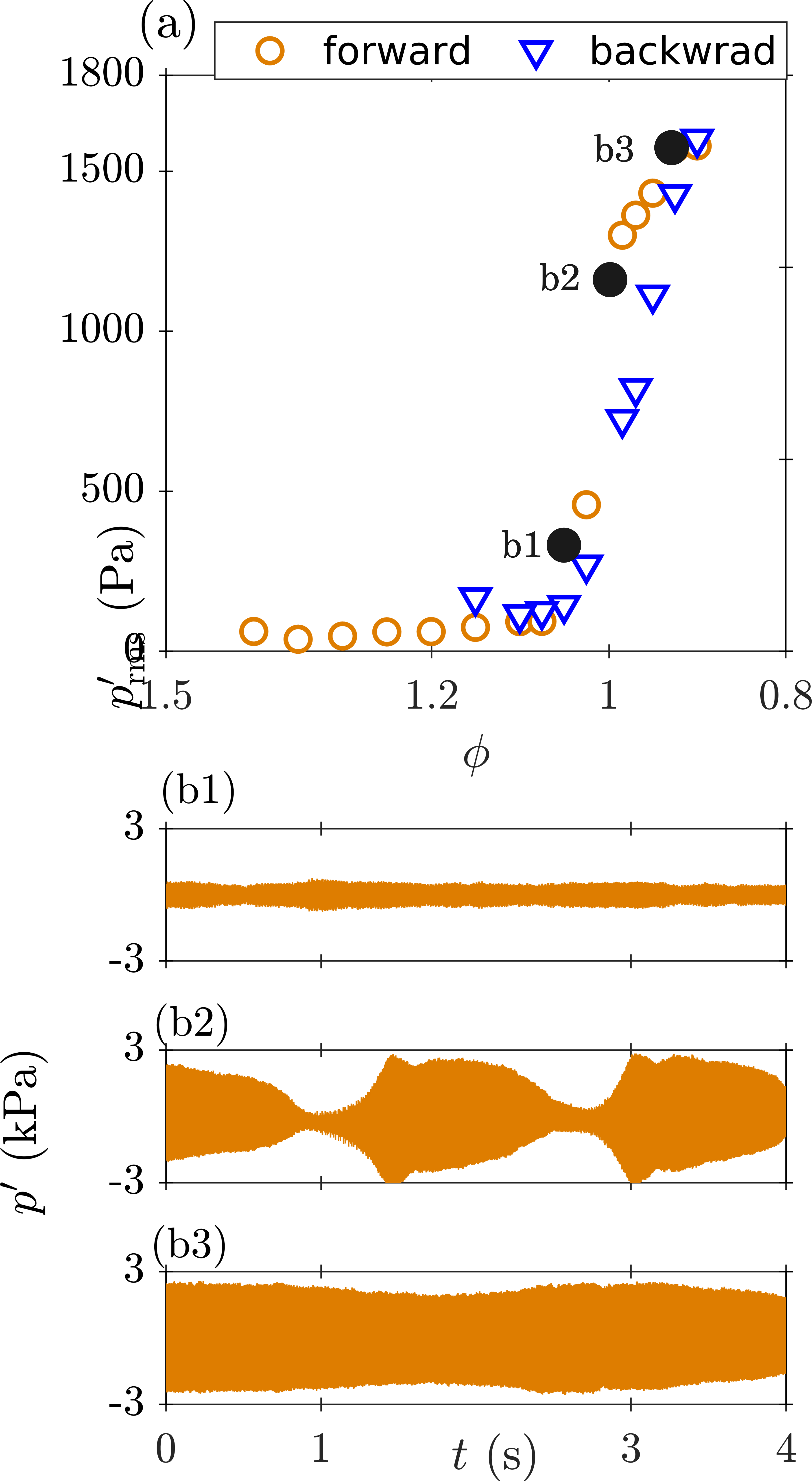}
\caption{Representation of the sudden transition via canard explosion in an annular combustor. \textbf{(a)} The forward and reverse path of the transition via the canard explosion. The equivalence ratio ($\phi$) in the forward path is varied from 1.4 to 0.9. We notice that the transition occurs via the state of large amplitude bursting \textbf{(b1-b3)}.}
    \label{Fig: Bifurcation annular}
\end{figure}

Further, in the swirl stabilized dump combustor, the steep rise in RMS value of $p'$ to a high amplitude oscillatory instability occurs via the state of large amplitude bursting (refer to point a2 in Fig.~\ref{Fig: Bifurcation swirler}a). The state of bursting has imprints corresponding to the states of low-amplitude fluctuations and high-amplitude fluctuations (Fig.~\ref{Fig: Bifurcation swirler}a2 \& a1). Similarly, when $\phi$ is varied from 1.4 to 0.9 in an annular combustor, we observe a sudden transition for $\phi > 1.075$ (refer to Fig.~\ref{Fig: Bifurcation annular}a). Upon varying $\phi$ in the reverse direction, the transition retraces its path. Moreover, we note that the transition occurs via a state of large amplitude bursting (refer to Fig.~\ref{Fig: Bifurcation annular}b2). The bursting state has the imprints of low-amplitude fluctuations and high-amplitude fluctuations (cf. ref Fig.~\ref{Fig: Bifurcation annular}b1, b2 \& b3), similar to the bursting characteristics observed in the swirl stabilized dump combustor. However, the time interval of bursting oscillations in the annular combustor is larger than the time interval of bursting in the swirl-stabilized combustor. In summary, in all three combustors, we observe that the amplitude of the bursts corresponding to an underlying canard explosion is very high due to the rapid nature of the transition at the bifurcation regime.

In order to investigate the bursting phenomenon, we experimentally measure temperature fluctuations of the hot exhaust gases for the swirl-stabilized dump combustor during the state of bursting ($\phi = 0.657$). The temperature fluctuations are measured using a K-type thermocouple. The exhaust gas temperature is governed by the internal variables of the combustor, such as flame temperature, equivalence ratio and the heat transfer rate to the combustor walls. These variables, in turn, govern the dynamics of the oscillatory instabilities exhibited by a combustor.

\begin{figure}
    \centering
    \includegraphics[width=0.6\linewidth]{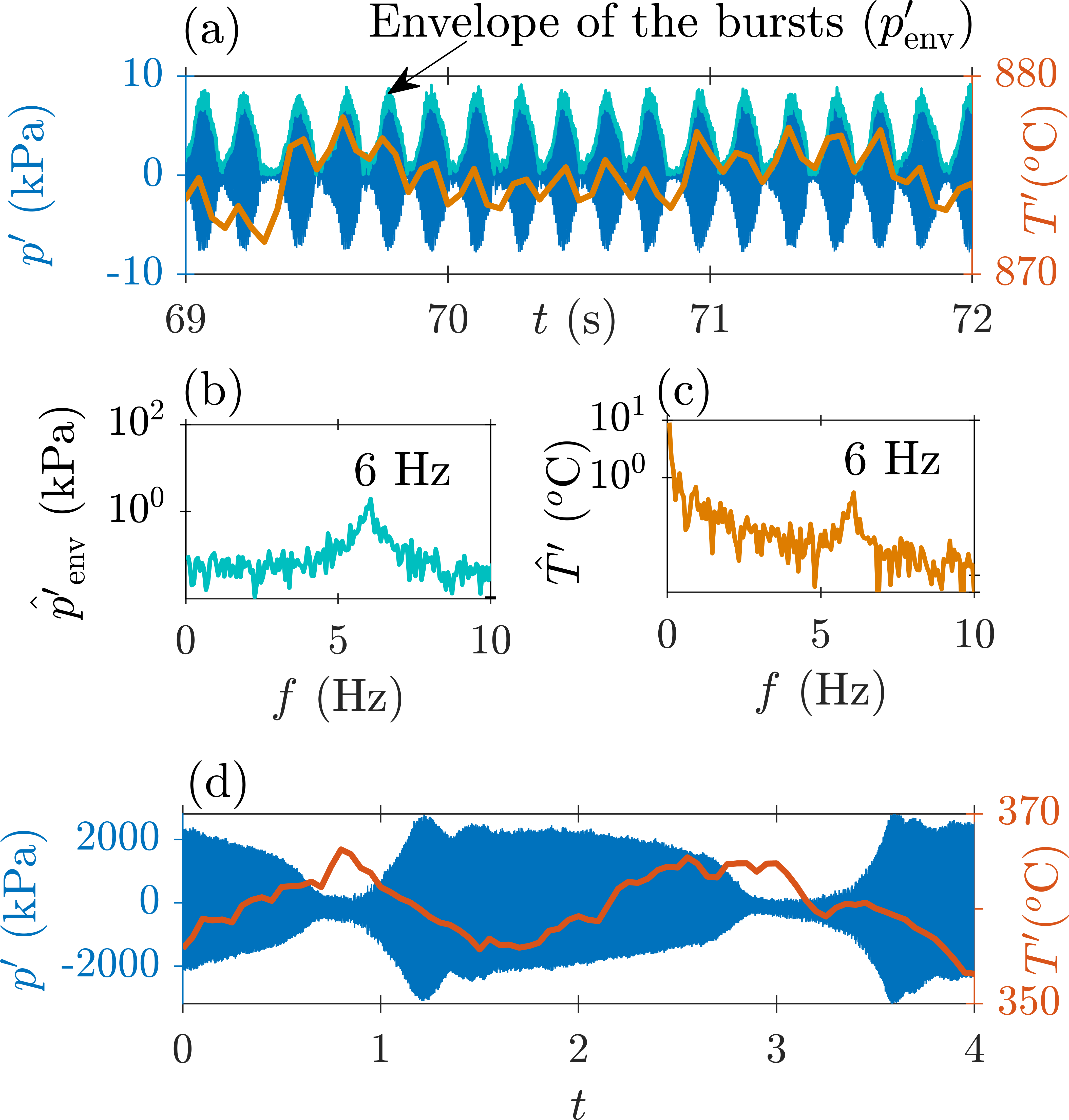}
    \caption{(\textbf{a}) Representation of the variation of the exhaust gas temperature ($T'$) along with the acoustic pressure fluctuations ($p'$) for a swirl stabilized dump combustor, measured during the state of bursting via canard explosions ($\phi = 0.657$). (\textbf{b-c}) The amplitude spectrum of the envelope of acoustic pressure fluctuations ($\hat{p}'_\mathrm{env}$) and the temperature fluctuations ($\hat{T}'_\mathrm{env}$) have the same dominant frequency at 6 Hz. (\textbf{d}) The variation of $T'$ along with $p'$ for an annular combustor, measured during the state of bursting via canard explosions ($\phi = 1$). Notice the pattern of variations in $T'$ and the envelope of the $p'$; the maxima of $T'$ is in the region of minimum $p'$, and the minima of $T'$ is in the region of maximum $p'$.}
    \label{Fig: Temperature variation with pressure}
\end{figure}

Figure~\ref{Fig: Temperature variation with pressure} represents the variation in temperature alongside the acoustic pressure fluctuation $p'$ during bursting in a swirl stabilized dump combustor. We note that there is a strong correlation between the temperature fluctuation ($T'$) and the envelope of the bursting oscillations ($p'_\mathrm{env}$). The strength of the correlation is tested by computing Pearson's correlation coefficient ($r$), and the value of $r$ is 0.84 for $T'$ and $p'_\mathrm{env}$. The time series of $T'$ is band passed to remove the fluctuations lesser than 1 Hz for computing the value of $r$. Moreover, the local maxima of $T'$ are in the high amplitude bursting regime of $p'$, and the local minima of $T'$ are in the low amplitude regime of $p'$ (Fig.~\ref{Fig: Temperature variation with pressure}a). This rhythmic variation of $T'$ and $p'_\mathrm{env}$ is also evident in the amplitude spectrum of the envelope of acoustic pressure fluctuations ($\hat{p}'_\mathrm{env}$) and the temperature fluctuations ($\hat{T}'_\mathrm{env}$) having the same dominant frequency at 6 Hz (refer to Fig.~\ref{Fig: Temperature variation with pressure}b,c). A similar observation of variation in $T'$ and the envelope of $p'$, but out of phase pattern, is made for the state of large amplitude bursting in the annular combustor at $\phi = 1$ (refer to Fig.~\ref{Fig: Temperature variation with pressure}d). 

Further, the past literature on bursting dynamics suggests that bursting occurs when a system parameter fluctuates at a slower time scale at the bifurcation regime \cite{izhikevich2000neural,kasthuri2019bursting,tandon2020bursting}. Therefore, observing variation in temperature fluctuations in correlation with the bursting amplitude (Fig.~\ref{Fig: Temperature variation with pressure}), we note that a system parameter is fluctuating at a slower time scale at the bifurcation regime.

Thus, it is evident from Figs.~\ref{Fig: Bifurcation bluff body}, \ref{Fig: Bifurcation swirler} and \ref{Fig: Bifurcation annular} that sudden transitions via canard explosions occur in three different turbulent reactive flow systems. Despite differences in the nature of the flow fields and the flame acoustic interactions in these different turbulent combustor configurations, we observe a common transition via canard explosion. The observation of large amplitude bursts in the regime of bifurcation hints towards an underlying universal mechanism, which we illustrate in the following subsections using a low-order model for thermo-fluid systems. Motivated by these results, we consider a modified Van der Pol oscillator as illustrated by \citet{ananthkrishnan1998application} to describe the sudden transition. We reduce the influence of the lower-order nonlinearities such that the variation of the system amplitude becomes highly sensitive to the control parameter at the bifurcation regime. We further incorporate a slowly varying coupling term to the acoustic driving to obtain the phenomenon of large amplitude bursting. 

\section{Modelling canard explosion in thermoacoustic system}
\label{Sec: Thermoacoustic model of Canard explosio}
\begin{figure} 
    \centering
    \includegraphics[width =0.72\linewidth]{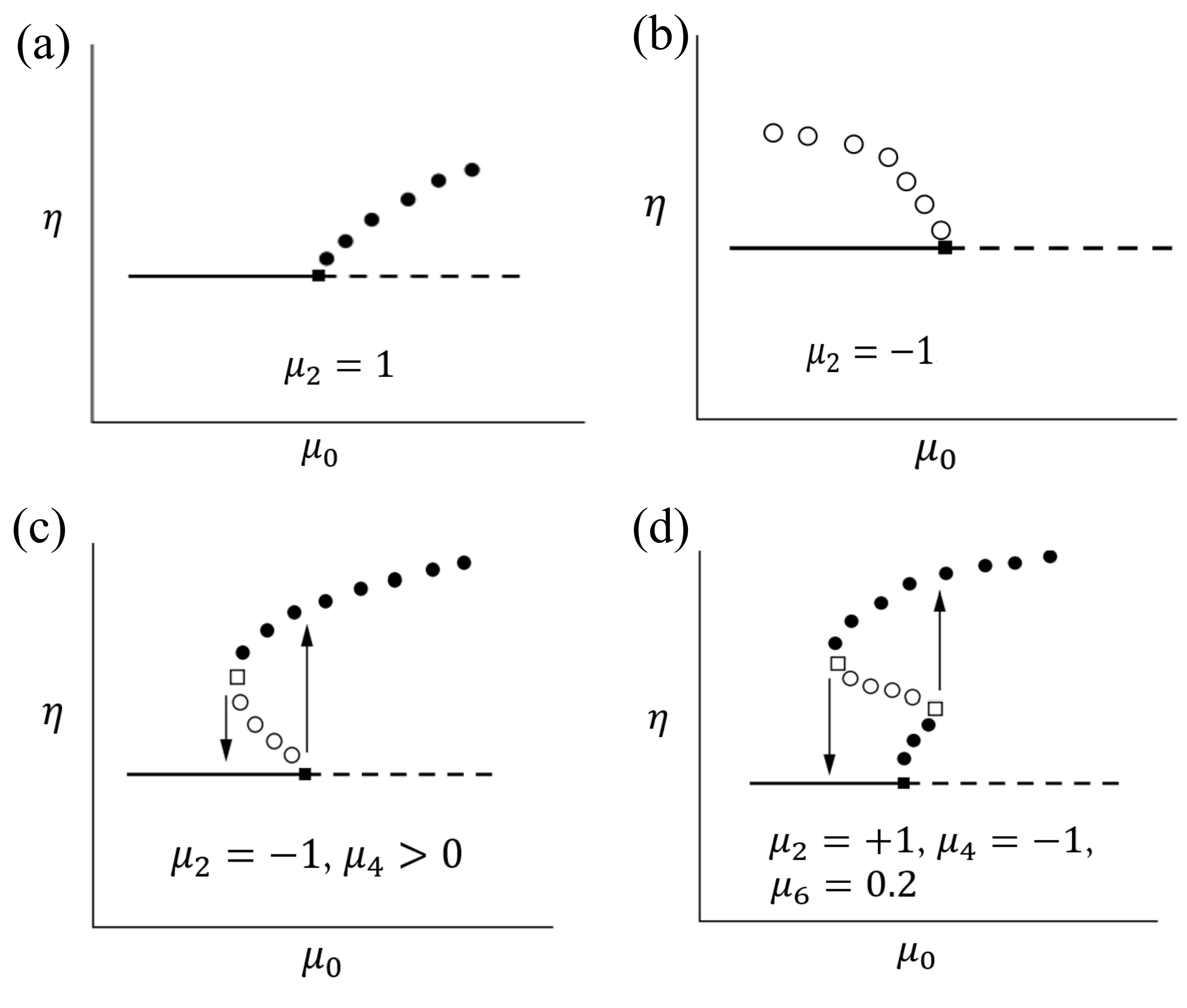}
    \caption{Representation of the types of bifurcation obtained by augmenting the driving term $\dot{\eta}$ of Eq.~(\ref{Eq: Modified Van der Pol equation}) with the higher order nonlinear terms. \textbf{(a)} Supercritical Hopf bifurcation with a single stable branch of LCO. \textbf{(b)} Subcritical Hopf bifurcation with a single unstable LCO branch. \textbf{(c)} Subcritical Hopf bifurcation to a stable LCO branch. \textbf{(d)} Secondary bifurcation depicting a supercritical followed by an abrupt secondary transition to a high amplitude stable LCO. Open circles represent the unstable solutions, and the solid circles represent the stable solutions. This figure is reproduced with permission from \citet{ananthkrishnan1998application} }
    \label{Fig: explaining multiple branches}
\end{figure}

The dynamics of the canard explosion presented in the above experiments is mainly associated with the change in the amplitude of the acoustic pressure fluctuations as the parameter is varied. Since we are concerned with modelling the rapid continuous rise in the amplitude, the thermoacoustic system considered here is one-dimensional, where the axial modes are excited. The effects of mean flow and temperature gradient are neglected \cite{nicoud2009zero,balasubramanian2008thermoacoustic}. The nonlinear acoustic terms are considered insignificant as the pressure fluctuations relative to the mean are negligible. Thus, the dynamics of the acoustic pressure and the heat release rate fluctuations inside the combustion chamber is governed by the linearized momentum and energy conservation equations \cite{balasubramanian2008thermoacoustic}, which are given as,
\begin{align}
    \frac{1}{\bar\rho} \frac{\partial p^\prime(z,t) }{\partial  z} + \frac{\partial u^\prime(z,t) }{\partial  t} &= 0,\label{ET1}\\
    \frac{\partial p^\prime(z,t) }{\partial  t} + \gamma \bar p \frac{\partial u^\prime(z,t) }{\partial  z} &= (\gamma -1) \dot{ Q}^\prime(z,t) \delta( z-{z_f}).\label{ET2}
\end{align}
Here, $t$ is time, $z$ is the distance along the axial direction of the duct, and $\gamma$ is the specific heat ratio. $\bar{\rho}$ and $\bar{p}$ indicate the mean density and pressure, while $p^\prime$ and $u^\prime$ are the pressure and velocity fluctuations, respectively. We assume the chemical reaction zone to be of a smaller volume such that the heat release rate fluctuations  $\dot{Q}'$ are concentrated at a location $z_f$, which is represented by a Dirac-delta ($\delta$) function \cite{mcmanus1993review}. Equations (\ref{ET1}) and (\ref{ET2}) can be appropriately modified to obtain an inhomogeneous wave equation, as given below \cite{lieuwen2021unsteady}: 

\begin{equation}\label{ET5}
    \begin{aligned}
        c^2 \frac{\partial^2 p'(z,t)}{\partial z^2} - \frac{\partial^2 p'(z,t)}{\partial t^2} \\ 
        = -(\gamma -1) \frac{\partial \dot{ Q}'(z,t)}{\partial t} \delta( z-{z_f}),
    \end{aligned}
\end{equation}
where, $c = \sqrt{\gamma\bar{p}/\bar{\rho}}$ is the speed of sound. We obtain an ordinary differential equation by simplifying Eq.~(\ref{ET5}) using a Galerkin modal expansion \cite{lores1973nonlinear}. The $u'$ and $p'$ are projected on a set of spatial basis functions (sines and cosines). The temporal coefficients of the basis functions are $\eta$ and $\dot{\eta}$, and are represented as:
\begin{equation}\label{ET6}
    \begin{aligned}
        p'( z,t) &= \bar p \sum_{j = 1}^{n} \frac{\dot\eta_j(t)}{\omega_j} \cos(k_j z)  \quad\text{and}\quad \\
        u'( z,t) &= \frac{\bar p}{\bar\rho  c} \sum_{j = 1}^{n} \eta_j(t) \sin(k_j z),
    \end{aligned}
\end{equation}
where $j$ represent the eigenmodes. The basis functions satisfy the acoustic boundary conditions--- i.e., $u'=0$ at the closed end and $p'=0$ at the open end of the duct. The chosen basis functions are orthogonal in nature. These basis functions also form the eigenmodes of the self-adjoint part of the linearized equations \cite{balasubramanian2008thermoacoustic}. Here, for a given length of the combustor $L$, $k_j$ is the wavenumber ($k_j = (2j-1) \pi /2 L $). The wavenumber is related to the natural frequency as $\omega_j =  c k_j$. After substituting for Eq.~(\ref{ET6}),  Eq.~(\ref{ET5}) can be written as, 
\begin{equation}\label{ET7}
    \begin{aligned}
        \sum_{j = 1}^{n} \frac{\ddot \eta_j(t)}{\omega_j} \cos{(k_j z)}+\frac{\gamma \bar p}{\bar \rho  c} \sum_{j = 1}^{n} \eta_j(t) k_j \cos(k_j z) \\
        = \frac{\gamma - 1}{\bar p} \dot Q' \delta(z-z_f).
    \end{aligned}    
\end{equation}
By integrating Eq.~(\ref{ET7}) over the volume of the combustor, after computing the inner product along each of the basis functions, we obtain
\begin{equation}\label{ET8}
    \frac{\ddot\eta_j(t)}{\omega_j}+  c k_j \eta_j(t) = \frac{2(\gamma -1)}{L \bar p} \int_{0}^{L} \dot Q' \delta(z-z_f) \cos(k_j z) \mathrm {d}z.
\end{equation}
Here, we choose the number of eigenmodes to be $j = 1$, which is adequate for analysing the characteristics of the transition discovered in the experiments conducted in the current study. Further, the observed dynamics in the combustors is a result of nonlinear response of the flame to the fluctuations in the acoustic field. Therefore, $\dot{Q}'$ can be expressed as a nonlinear function of $\eta$ and $\dot\eta$. Thus, Eq.~(\ref{ET8}) reduces to the equation of a self-excited harmonic oscillator, expressed as
\begin{equation}\label{ET16}
    \ddot{\eta} + \omega^2 \eta = f(\eta, \dot{\eta}),
\end{equation}
where, $f(\eta,\dot \eta) = f(\dot Q^\prime) - \alpha \dot \eta$ is the nonlinear driving term. An extra term $\alpha \dot \eta$ is added to take acoustic damping into account ($\alpha$ is the damping coefficient) \cite{noiray2017linear}. Thus, the source term $f(\eta,\dot \eta)$ represents the nonlinear damping and driving behaviour of the oscillator. Further, $f(\eta,\dot \eta)$ can be expanded with nonlinear terms such that Eq.~(\ref{ET16}) represents a Hopf bifurcation to thermoacoustic oscillations \cite{bonciolini2021low}. The modified form of Eq.~(\ref{ET16}) is given as,
\begin{equation}\label{Eq: Basic Van der Pol equation}
    \ddot{\eta}+\left (\mu_2 \eta^2- \mu_0 \right ) \dot{\eta}+\omega^2 \eta=0,
\end{equation}
where $\mu_0$ is the control parameter and $\mu_2$ is the coefficient of the second order nonlinear term. Equation \ref{Eq: Basic Van der Pol equation} also represents the Van der Pol oscillator, which is a paradigm for systems exhibiting limit cycle oscillations \cite{minorsky1962nonlinear}. When $\mu_2$ is positive, we obtain a stable limit cycle branch denoting a supercritical Hopf bifurcation (refer to Fig.~\ref{Fig: explaining multiple branches}a). When $\mu_2$ is negative, we obtain an unstable subcritical limit cycle branch (refer to Fig.~\ref{Fig: explaining multiple branches}b).
The nonlinear coefficients associated with the driving term $\dot{\eta}$ in Eq.~(\ref{Eq: Basic Van der Pol equation}) can be augmented with higher order nonlinear coefficients to produce multiple limit cycle branches \cite{ananthkrishnan1998application}. This augmentation helps represent the multiple high amplitude limit cycle oscillations (LCO) in thermoacoustic systems \cite{bhavi2023abrupttransition}. Therefore, we modify Eq.~(\ref{Eq: Basic Van der Pol equation}) as,
\begin{equation}\label{Eq: Modified Van der Pol equation}
    \ddot{\eta}+\left (\mu_6 \eta^6+ \mu_4 \eta^4+\mu_2 \eta^2\right ) \dot{\eta}- \mu_0  \dot{\eta}+\omega^2 \eta = 0,
\end{equation}
where $\mu_4$ and $\mu_6$ are the coefficients of the higher order nonlinear terms. By fixing $\mu_2 = -1$, $\mu_4 > 0$ and $\mu_6 = 0$, we obtain an unstable LCO branch followed by a stable LCO branch representing a subcritical Hopf bifurcation (Fig.~\ref{Fig: explaining multiple branches}c). Similarly by fixing $\mu_2 > 0$, $\mu_4 < 0$ and $\mu_6 > 0$, we obtain a secondary bifurcation as shown in Fig.~\ref{Fig: explaining multiple branches}d \cite{ananthkrishnan1998application,bhavi2023abrupttransition}. Thus, from Fig.~\ref{Fig: explaining multiple branches}, we note that the coefficients of the nonlinear terms govern the stability and the amplitude of the LCO branches in the bifurcation curve.  
\begin{figure}[h]
    \centering
    \includegraphics[width=0.5\linewidth]{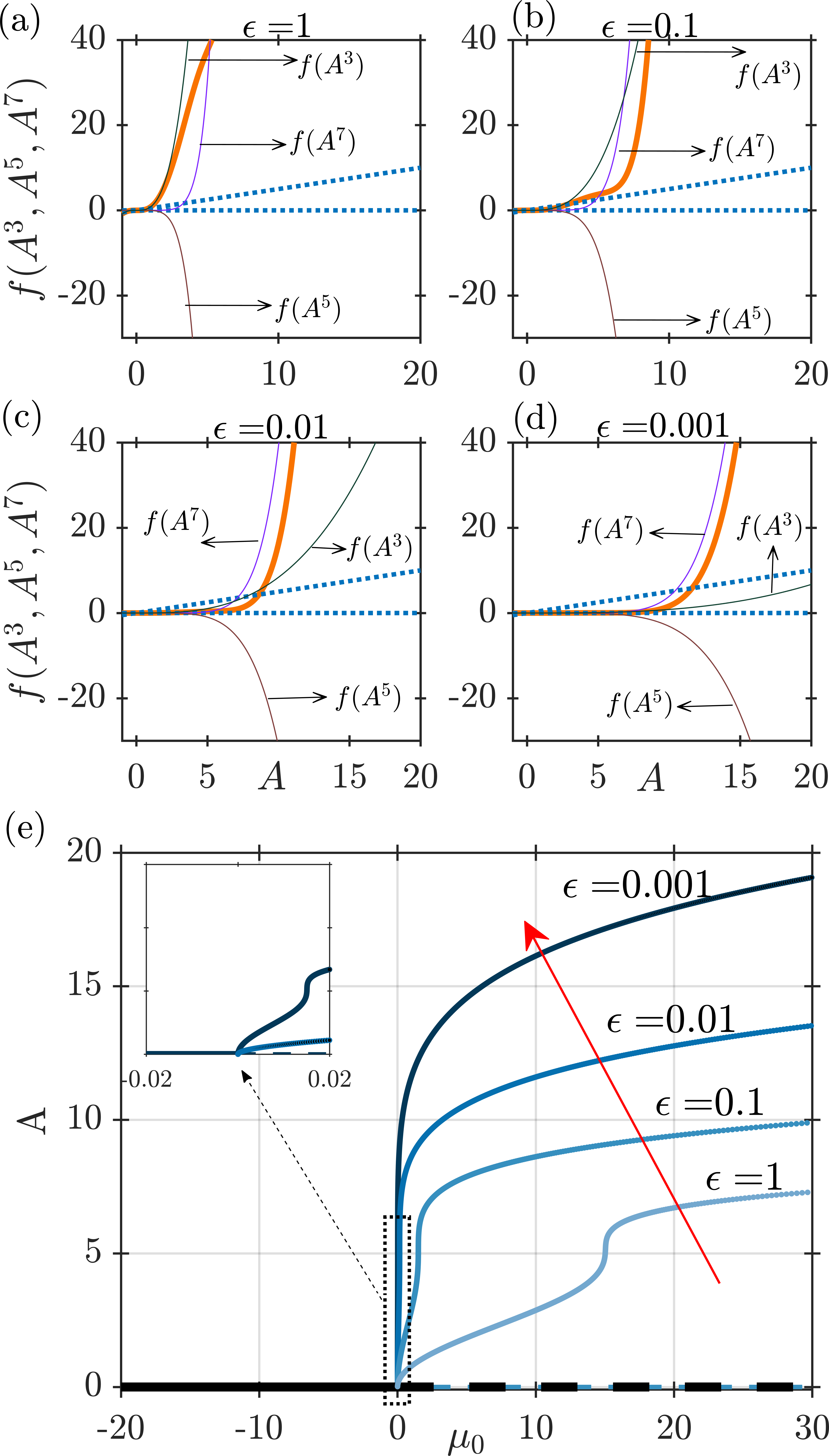}
    \caption{ \textbf{(a-d)} Representation of the effects of $\epsilon$ on the evolution of the solutions for Eq.~(\ref{Eq: Amplitude envelope}), which is of the form $\dot{A} = f(\mu_0,~A)- f(A^3,~A^5,~A^7)$. The ordinate denotes the values of $f(A^3,~A^5,~A^7)$ (thick orange curve), and the abscissa represents the values of $A$. The contributions from $A^3$, $A^5$, and $A^7$ are presented in the coloured thin solid curves. The dashed blue lines represent the control parameter curve $f(\mu_0,~A)$ for $\mu_0 = $ 0 and 1. Geometrically, the solutions are the points of intersections of $f(A^3,~A^5,~A^7)$ and $f(\mu_0,~A)$. \textbf{(e)} Bifurcation diagram to represent the effect of $\epsilon$ on the continuous secondary bifurcation curve obtained by fixing $\mu_2 = 6.7$, $\mu_4=-0.5$, and $\mu_6 = 0.01$. The variation of $\mu_0$ is shown on the abscissa and the solutions of Eq.~(\ref{Eq: Amplitude envelope}) are shown on the ordinate. In the bifurcation diagram, thick lines are for stable solutions and the broken line is for unstable solutions. Notice that as $\epsilon$ is decreased from 1 to 0.001, the range of parameters to reach the high amplitude oscillations after the bifurcation decreases to a very narrow span.}
    \label{Fig: illustration to obtain the canard explosion}
\end{figure}

Now, the dynamics of the canard explosion is such that the amplitude of the system becomes highly sensitive to a narrow range of parameters near the bifurcation regime. To achieve this, we reduce the magnitude of the coefficients of all the nonlinear terms $\left (\mu_6 \eta^6+ \mu_4 \eta^4+\mu_2 \eta^2\right )$. Therefore, we couple all the nonlinear coefficients with a constant $\epsilon \ll 1$, reducing the strength of nonlinearity associated with the nonlinear terms. Such systems with reduced strength of nonlinearity are referred to as weakly nonlinear oscillators \cite{strogatz2018nonlinear}. The modified equation with the coupling term $\epsilon$ is written as,

\begin{equation}\label{Eq: Canard Van der Pol equation}
    \ddot{\eta}+ \epsilon \left (\mu_6 \eta^6+ \mu_4 \eta^4+\mu_2 \eta^2\right ) \dot{\eta}- \mu_0  \dot{\eta}+\omega^2 \eta = 0.
\end{equation} 

To visualise the effect of the magnitude of $\epsilon$, we obtain the dynamics of the amplitude-envelope of the oscillations from the harmonic oscillator Eq.~(\ref{Eq: Canard Van der Pol equation}), using the method of averaging \cite{bhavi2023abrupttransition,balanov2009simple,strogatz2018nonlinear}. We substitute the acoustic variable to be of the form $\eta(t) = A(t) cos[\omega t + \Omega(t)]$. Here, $A(t)$ and $\Omega(t)$ represent the amplitude-envelope and its phase, respectively. The evolution time scale of $A(t)$ and $\Omega (t)$ is much slower than the faster times scale of system $2\pi/\omega$. Thus, after substituting $\eta(A,\Omega)$ and averaging Eq.~(\ref{Eq: Canard Van der Pol equation}) over the faster time scale $2\pi/\omega$ \cite{balanov2009simple,bhavi2023abrupttransition}, the dynamics of the amplitude-envelope of the oscillations is obtained as, 
\begin{equation}\label{Eq: Amplitude envelope}
    \dot{A} = \frac{\mu_0}{2} A - \epsilon \left ( \frac{\mu_2}{8} A^3+ \frac{\mu_4}{165} A^5+ \frac{5 \mu_6}{128} A^7\right ).
\end{equation}
We note that the evolution of the amplitude-envelope is a function $\dot{A} = f(\mu_0,~A)- f(A^3,~A^5,~A^7)$, which is dependent on the control parameter $\mu_0$ and the damping term  $f(A^3,~A^5,~A^7)$. The nonlinear damping term $f(A^3,~A^5,~A^7)$ is in turn a function of the higher order terms $f(A^3)$, $f(A^5)$ and $f(A^7)$. The solutions for Eq.~(\ref{Eq: Amplitude envelope}) are computed as $\dot{A} = 0$, which are obtained by balancing $f(\mu_0,~A) = f(A^3,~A^5,~A^7)$ \cite{strogatz2018nonlinear}.

We proceed with considering a case of continuous secondary bifurcation obtained by setting $\mu_2 = 6.7$, $\mu_4 = -0.5$, and $\mu_6 = 0.01$. In Fig.~\ref{Fig: illustration to obtain the canard explosion}(a-d), we represent the effect of $\epsilon$ on the evolution of solutions for Eq.~(\ref{Eq: Amplitude envelope}). These solutions are, geometrically, the points of intersections of the curves $f(\mu_0,~A)$ and $f(A^3,~A^5,~A^7)$. The thick orange line represents the curves for $f(A^3,~A^5,~A^7)$, which is a summation of contributions from $f(A^3)$, $f(A^5)$ and $f(A^7)$ represented with thin lines. The curves for $f(\mu_0,~A)$, at $\mu_0 = 0$ and $\mu_0 = 1$, are shown in dotted blue lines. $f(\mu_0,~A)$ is a line passing through the origin where $\mu_0$ is its slope. Thus, we obtain several curves for $f(\mu_0,~A)$ with varying slopes as we vary $\mu_0$ as a control parameter, not shown here in the interest of space. From the figure~\ref{Fig: illustration to obtain the canard explosion}(a,b), for the lower values of $A$, we see that the dynamics of the curve $f(A^3,~A^5,~A^7)$ (orange line) is mainly contributed from $f(A^3)$ and $f(A^5)$. We also note that as the value of $\epsilon$ decreases from 1 to 0.001, the absolute value of the functions ($|f(A^3)|$,$|f(A^5)|$, and $|f(A^7)|$) decreases, and their curves tend towards the abscissa (cf. Fig.~\ref{Fig: illustration to obtain the canard explosion}a-d). The effect of the decrease in $\epsilon$, for smaller amplitudes of $A$, is more pronounced on the lower order nonlinear terms $f(A^3)$ and $f(A^5)$ than on the highest order term $f(A^7)$ (cf. Fig.~\ref{Fig: illustration to obtain the canard explosion}a-d). This influence of $\epsilon$ on the nonlinear terms collectively transforms the curve $f(A^3,~A^5,~A^7)$ to have lower slopes for an extended value of $A$ (compare the orange lines of Fig.~\ref{Fig: illustration to obtain the canard explosion}a-d). Thus, the transformation results in a scenario where we observe a rapid change in the value of solutions, the intersection of $f(A^3,~A^5,~A^7)$ and $f(\mu_0,~A)$ (cf. Fig.~\ref{Fig: illustration to obtain the canard explosion}c,d), for a minute change in the value of the parameter $\mu_0$ in the range $|\mu_0| < 1$.

In Fig.~\ref{Fig: illustration to obtain the canard explosion}(e), we plot the bifurcation curves for the cases of $\epsilon$ = 1, 0.1, 0.01, and 0.001 obtained by varying the control parameters in the range of $ -20 \le \mu_0 \le 30 $. As $\epsilon$ is reduced, we notice that the bifurcation curve significantly steepens at the Hopf point $\mu_0 = 0$ (refer to Fig.~\ref{Fig: illustration to obtain the canard explosion}e). In other words, the range of values of $\mu_0$ to reach the saturation in the rise in amplitude decreases to a very narrow span (refer to the inset of Fig.~\ref{Fig: illustration to obtain the canard explosion}e). The steepening of the transition curve occurs due to the higher reduction in the nonlinearity of lower-order nonlinear terms for lower amplitudes $A$, which otherwise form a continuous secondary bifurcation (refer to the curve $\epsilon = 1$, in Fig.~\ref{Fig: illustration to obtain the canard explosion}). From figure.~\ref{Fig: illustration to obtain the canard explosion}(a-d), we convey that the effect of $\epsilon$ is less on the highest-order nonlinear term $\eta^6$ when compared to the lower-order nonlinear terms in Eq.~(\ref{Eq: Canard Van der Pol equation}). Thus, the coupling term $\epsilon$ aids in obtaining a weakly nonlinear oscillator exhibiting a transition with a canard explosion at the Hopf point.

\begin{figure} 
    
    \centering
    \includegraphics[width=0.6\linewidth]{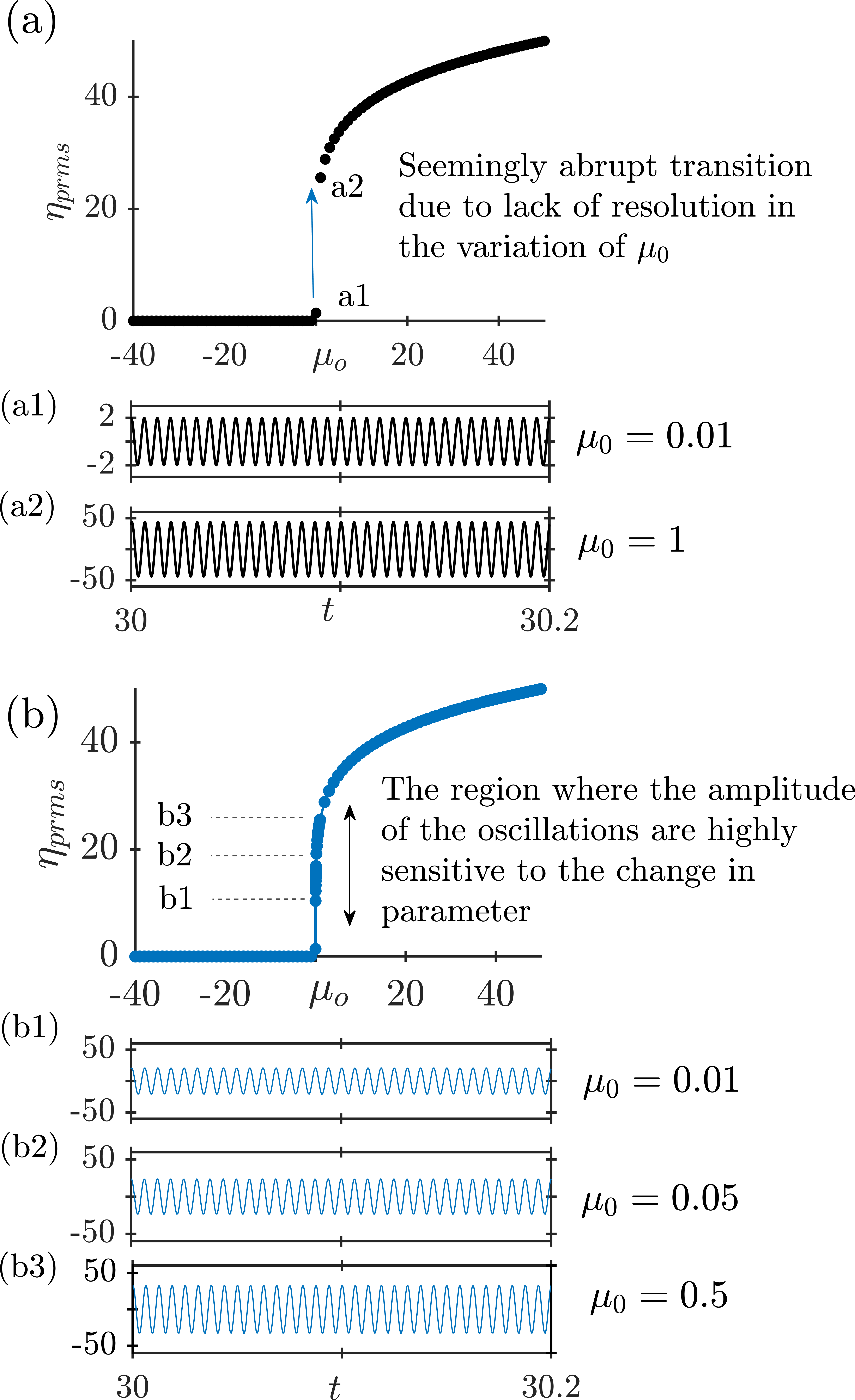}
    \caption{Representation of the canard explosion by numerically integrating Eq.~(\ref{Eq: Canard Van der Pol equation}). The curves represent the variation of $\eta_{rms}$ as a function of $\mu_o$. When the control parameter $\mu_o$ is varied in steps of 1, we notice an abrupt jump in \textbf{(a)} the bifurcation diagram. The abrupt nature of the transition is due to a lack of resolution in the variation of the control parameter. The abrupt jump is also evident from the amplitude of the time series \textbf{(a1)} before and \textbf{(a2)} after the transition. However, when we vary the control parameter in finer steps, we have \textbf{(b)} stable dynamics at each of these finer steps. Thus, the model captures a rapid continuous transition \textbf{(b1-b3)}, where the amplitude rises significantly with a negligible change in the control parameter $\mu_0$.}
    \label{Fig: model visualisation of canard explosions} 

\end{figure}

Further, utilising $4^\mathrm{th}$ order Runge-Kutta method, we numerically integrate Eq.~(\ref{Eq: Canard Van der Pol equation}) by fixing $\epsilon = 0.0001$ for a range of control parameter $-40 \le \mu_0 \le 50$ to obtain the bifurcation diagram. Figure~\ref{Fig: model visualisation of canard explosions}a denotes the bifurcation curve when the control parameter $\mu_o$ is varied in steps of 1. Since there is a significantly steeper rise, the transition appears to be abrupt at the Hopf point $\mu_0 = 0$ due to a weaker resolution in the variation of the control parameter. This seemingly abrupt transition is what we notice during the experiments as the system transitions to high-amplitude thermoacoustic instability (refer to Fig.~\ref{Fig: Bifurcation bluff body}a). We further illustrate that, by increasing the resolution at the canard explosion regime, the system exhibits the stable LCO at every small variation in $\mu_0$, implying a continuous sudden transition (refer to Fig.~\ref{Fig: model visualisation of canard explosions}b).

Further, the experimental data on temperature fluctuations vary in correlation with the bursting amplitude at a slower time scale (refer to Fig.~\ref{Fig: Temperature variation with pressure}). This variation of the temperature fluctuations suggests that there is an additional parameter that fluctuates at a timescale slower than the thermoacoustic oscillations. When such an oscillating term is coupled with the driving term $\dot{\eta}$, the system exhibits bursting oscillations at the bifurcation regime \cite{kasthuri2019bursting}. We illustrate the bursting phenomenon for an underlying canard explosion in the following subsection.

\subsection{Bursting behaviour due to underlying canard explosion}\label{Sec: Illustrate bursts using model}
The amplitude of the bursts corresponding to an underlying canard explosion is very high due to the sudden nature of the transition at the bifurcation regime. Experimentally, we observed that a system parameter ($T'$) fluctuates at a slower time scale at the bifurcation regime of the canard explosion (refer to Fig.~\ref{Fig: Temperature variation with pressure}). Such parametric oscillations are also reported in past studies of thermoacoustic systems \cite{kasthuri2019bursting,tandon2020bursting}. \citet{kasthuri2019bursting} showed that the temperature close to the burner oscillates at a much slower time scale than the thermoacoustic oscillations during the state of bursting. In a swirl stabilized turbulent combustor, \citet{hong2008generation} showed that there is a fluctuation in the equivalence ratio during the state of large amplitude bursting. \citet{tandon2020bursting} replicated the bursting dynamics of the low-turbulence systems using a phenomenological model containing slow-fast time scales. In line with the conjectures of these studies, one would intuitively expect large amplitude bursting oscillations in a system containing slow-fast time scales across the canard explosions. Inspired by these studies, we further illustrate the effect of the fluctuation of the system parameter at the bifurcation regime of a canard explosion; for that, we couple the driving term $\dot{\eta}$ with a periodic oscillation of a very low frequency $\omega_q$ with a coupling strength of $q$. Thus, Eq.~(\ref{Eq: Canard Van der Pol equation}) is further modified as,
\begin{equation}\label{Eq: Canard model with noise}
    \begin{aligned}
        \ddot{\eta}+ \epsilon \left (\mu_6 \eta^6+ \mu_4 \eta^4+\mu_2 \eta^2\right ) \dot{\eta} - \mu_0  \dot{\eta} \\ 
       -[q \sin (\omega_q t) + \xi_m] \dot{\eta} +\omega^2 \eta + \xi_a = 0.
    \end{aligned}
\end{equation}
The coupling is added with the multiplicative noise $\xi_m$ to model the fluctuations associated with the driving as a result of the internal noise in the system \cite{clavin1994turbulence}. We also add additive white noise $\xi_a$ to the Eq.~(\ref{Eq: Canard model with noise}) to incorporate the effect of turbulence \cite{noiray2017linear}. Here, $\xi$ is the white noise defined as $\langle \xi\xi_{\tau} \rangle = \Gamma \delta{\tau}$, where $\Gamma$ is the noise intensity. The subscripts `$m$' and `$a$' denote the correspondence to multiplicative and additive noise, respectively.

\begin{figure} [h]
    \centering
    \includegraphics[width=0.6\linewidth]{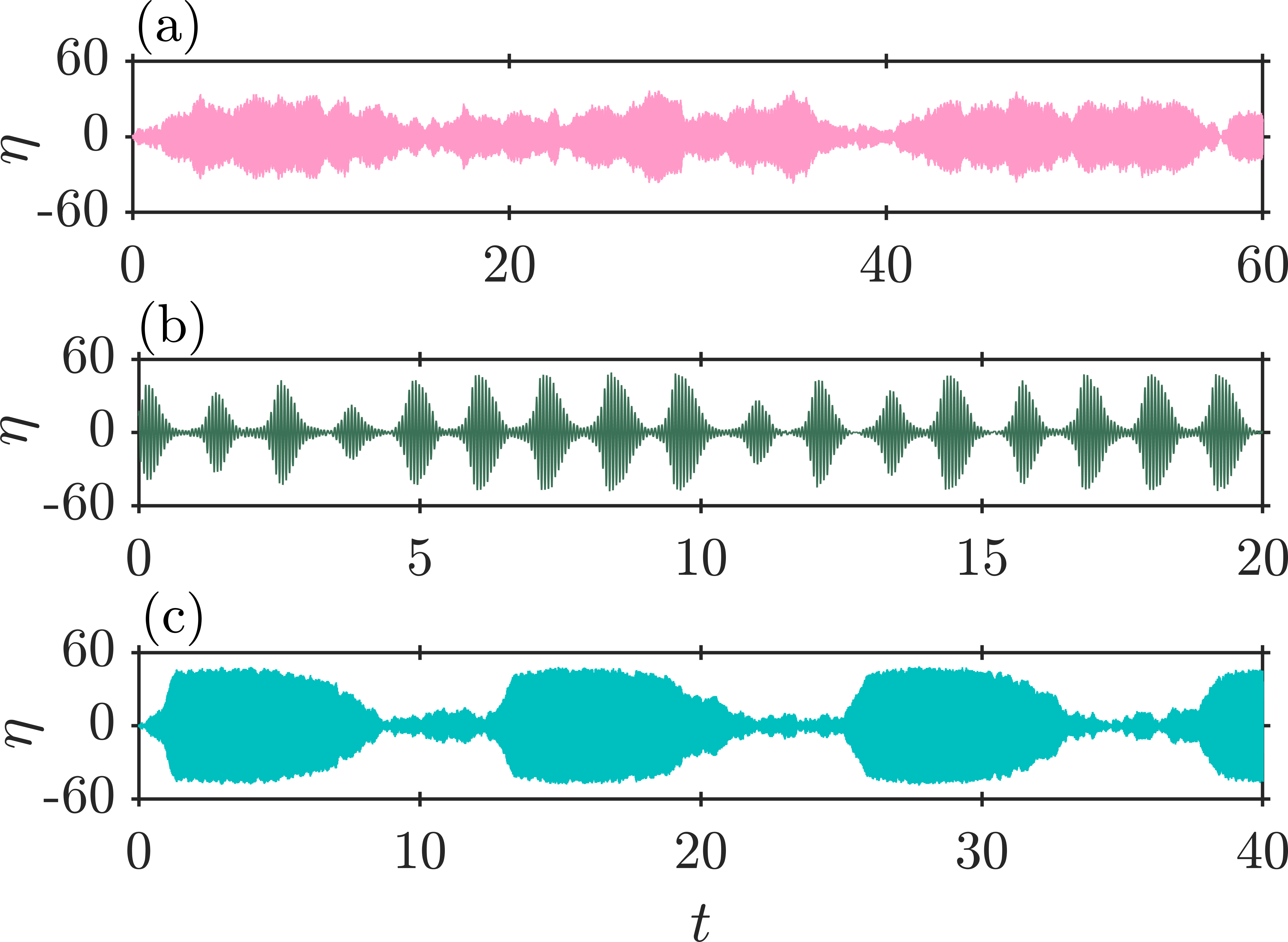}
    \caption{Representation of the time series for bursting behaviour at the bifurcation regime of the canard explosion, obtained by numerical integration of Eq.~(\ref{Eq: Canard model with noise}). \textbf{(a)} Time series analogous to the bursting behaviour of a bluff body stabilized dump combustor. \textbf{(b)} Time series analogous to the bursting behaviour of a swirl stabilized dump combustor. \textbf{(c)} Time series analogous to the bursting behaviour of an annular combustor. We notice large amplitude bursts at the bifurcation regime due to the underlying canard explosion}
    \label{Fig: model time series for bursting behaviour} 
\end{figure}

The qualitative nature of the bursting behaviour obtained from the model for different types of combustors is represented in Fig.~\ref{Fig: model time series for bursting behaviour}. At $\mu_0 = 0$, fixing  $q = 0$, $\Gamma_a = 10^5$ and $\Gamma_m = 10^4$, we obtain the bursting behaviour that matches with the time series obtained from the bluff body stabilized dump combustor (refer to Fig.~\ref{Fig: model time series for bursting behaviour}a). The irregularity in the bursting pattern is due to the multiplicative noise $\xi_m$ associated with the driving term $\dot{\eta}$. When we fix $\omega = 370$ rad/s, $\omega_q = 3$ rad/s, $q = 20$ rad/s, $\Gamma_a = 10^5$ and $\Gamma_m = 10^4$, we obtain a bursting pattern observed in the swirl stabilized dump combustor (refer to Fig.~\ref{Fig: model time series for bursting behaviour}b).

Further, upon fixing $\omega = 370$, rad/s, $\omega_q = 0.5$ rad/s, $q = 20$ rad/s, $\Gamma_a = 10^7$ and $\Gamma_m = 10^5$, we obtain a bursting pattern observed in the annular combustor (refer to Fig.~\ref{Fig: model time series for bursting behaviour}b). The coupling oscillation frequency $\omega_q$ for the case of an annular combustor is lesser than that of the swirler stabilized dump combustor. Hence, the bursts in the annular combustor are of longer duration. Thus, using these results from the model we illustrate that large amplitude bursts are observed in turbulent combustors when a system parameter fluctuates at the bifurcation regime of an underlying canard explosion.

\section{Conclusions} \label{Sec: Conclusions}

In summary, we reported the experimental evidence for the occurrence of canard explosion in three different turbulent reactive flow systems---a bluff body and a swirl-stabilized dump combustor, and a swirl-stabilized annular combustor. The transition appears discontinuous when there is a lack of resolution in the variation of the control parameter. Though the rise in amplitude of the oscillations is steep in nature, unlike abrupt transitions, the canard explosion in this study exhibits no hysteresis. When such a transition involves a parameter fluctuation at the bifurcation regime, the system is bound to exhibit bursting behaviour with large amplitude bursts. We experimentally showed that the state of the bursting, in the regime of canard explosions, consists of very high amplitude fluctuations amidst low amplitude fluctuations.

We describe the transition via the canard explosion using the low-order model representing thermoacoustic systems. A continuous secondary bifurcation steepens at the bifurcation regime when the nonlinearity of the nonlinear damping in the model is reduced by coupling a small variable $\epsilon$. In other words, the dynamics of the transition from stable operation to high amplitude oscillatory instability gets restricted to a very narrow range of control parameters, for the values of $\epsilon \ll 1$. For such a steepened transition, we conjecture that the system amplitude becomes highly sensitive to the change in control parameter at the bifurcation regime, thus giving rise to a scenario of large amplitude bursts.

Further, during the state of bursting, we observe a slow variation in the fluctuation of the exhaust gas temperature in correlation with the envelope of the acoustic pressure fluctuation. The temperature of the exhaust gas represents the flame temperature as well as the fluctuation in the heat release rate, which in turn governs the dynamics of the thermoacoustic oscillations. We convey that parameter fluctuation has a role in bursting behaviour in the regime of canard explosion, as explained using the low-order thermoacoustic model. A further study that consists of flow visualisation is required to differentiate the underlying flow physics between the canard explosions, abrupt transition and gradual bifurcation in turbulent thermoacoustic systems. 
    
\begin{acknowledgments}
We acknowledge the support from Mr Pruthiraj M., Mr Rohit R. and Mr Beeraiah T. for their useful discussions while conducting experiments. We also thank Ms Athira, Ms Ariakutty, Mr Thilagaraj S. and Mr Anand S. for their technical support in experiments. Ramesh S. Bhavi and Sivakumar S. are thankful to the Ministry of Education (MoE) for the research assistantship. R. I. Sujith thanks the IoE initiative (SP22231222CPETWOCTSHOC) and SERB/CRG/2020/003051 from the Department of Science and Technology for funding this work.
\end{acknowledgments}

\section*{Data Availability}

The data that support the findings of this study are available from the corresponding author upon reasonable request.

\bibliography{Main}

\end{document}